\let\latexarabic\arabic
\let\latexdocument\document
\let\latexenddocument\enddocument

\RequirePackage[thmmarks]{ntheorem}
\makeatletter
\renewtheoremstyle{plain} 
  {\item[\hskip\labelsep \theorem@headerfont ##1\ \textup{##2}\theorem@separator]} 
  {\item[\hskip\labelsep \theorem@headerfont ##1\ \textup{##2}\ (##3)\theorem@separator]}
\makeatother

\documentclass[lineno]{biometrika}  

\let\document\latexdocument
\let\enddocument\latexenddocument
\AtEndDocument{\printhistory}
\let\arabic\latexarabic
\def\rm{}

\usepackage{amsmath}

\usepackage{times}
\usepackage{bm}
\usepackage{natbib}
\usepackage{graphicx, caption} 
\usepackage{multirow}
\usepackage{enumerate}
\usepackage{enumitem}
\usepackage{url}
\usepackage{subfiles}
\usepackage[caption = false]{subfig} 
\usepackage{color}

\usepackage{pifont}

\usepackage[plain,noend]{algorithm2e}

\begin{document}

\nolinenumbers

\markboth{H. Song and H. Chen}{Generalized kernel two-sample tests}

\title{Generalized kernel two-sample tests}

\author{Hoseung Song and Hao Chen}
\affil{Department of Statistics, University of California, Davis, Davis, California 95616, U.S.A.
\email{hosong@ucdavis.edu \ \  hxchen@ucdavis.edu}}

\def\by{y} 

\def\pr{\textsf{pr}}
\def\ep{\textsf{E}}
\def\var{\textsf{var}}       
\def\cov{\textsf{cov}}

\maketitle

\begin{abstract}
Kernel two-sample tests have been widely used for multivariate data to test equality of distributions. However, existing tests based on mapping distributions into a reproducing kernel Hilbert space mainly target specific alternatives and do not work well for some scenarios when the dimension of the data is moderate to high due to the curse of dimensionality. We propose a new test statistic that makes use of a common pattern under moderate and high dimensions and achieves substantial power improvements over existing kernel two-sample tests for a wide range of alternatives. We also propose alternative testing procedures that maintain high power with low computational cost, offering easy off-the-shelf tools for large datasets. The new approaches are compared to other state-of-the-art tests under various settings and show good performance. We showcase the new approaches through two applications: The comparison of musks and non-musks using the shape of molecules, and the comparison of taxi trips starting from John F. Kennedy airport in consecutive months. All proposed methods are implemented in an $\texttt{R}$ package $\texttt{kerTests}$.
\end{abstract}


\allowdisplaybreaks


\section{Introduction} \label{sec:int}


\subsection{Background} \label{sec:background}

Nonparametric two-sample hypothesis testing has received significant attention due to its relevance in many fields, often with the need to compare complex, high dimensional or large, data. Formally speaking, given two independent samples $X_{1}, X_{2}, \ldots, X_{m} \stackrel{iid}{\sim} P$ and $Y_{1}, Y_{2}, \ldots, Y_{n} \stackrel{iid}{\sim} Q$ where $P$ and $Q$ are distributions in $\mathcal{R}^{d}$, one wants to test $H_{0}: P = Q$ against $H_{1}: P \ne Q$. When $d$ is large, such as of the order of hundreds, thousands or more, it is common that one has little or no clue of  $P$ or $Q$, which makes parametric tests unrealistic in many applications. Several nonparametric tests have been proposed for high-dimensional data, including rank-based tests \citep{baumgartner1998nonparametric,hettmansperger1998affine,rousson2002distribution,oja2010multivariate}, inter-point distances-based tests \citep{szekely2013energy,biswas2014nonparametric,li2018asymptotic}, graph-based tests \citep{friedman1979multivariate,schilling1986multivariate,henze1988multivariate,rosenbaum2005exact,chen2017new},  and kernel-based tests \citep{gretton2007kernel,eric2008testing,gretton2009fast,gretton2012kernel, li2019optimality, gao2021two}. They all have succeeded in many applications. This work focuses on kernel-based tests.

The most well-known kernel two-sample test was proposed by \cite{gretton2007kernel}. In this framework, a reproducing kernel Hilbert space (RKHS) denoted as $H$ is employed, which is equipped with a continuous feature mapping $\phi(\cdot)$.  The inner product between feature mappings is defined by the positive definite kernel  function $k(X,Y) := <\phi(X), \phi(Y)>_H$.
The authors considered the maximum mean discrepancy (MMD) between two probability distributions $P$ and $Q$,
$\textrm{MMD}^2(P,Q) = E_{X,X'}[k(X,X')] - 2E_{X,Y}[k(X,Y)] + E_{Y,Y'}[k(Y,Y')]$,
where $X$ and $X'$ are independent random variables drawn from $P$, and $Y$ and $Y'$ are independent random variables drawn from $Q$. \cite{gretton2007kernel} considered two empirical estimates of $\textrm{MMD}^2(P,Q)$:
\begin{align}
\textrm{MMD}^2_{u} &= \frac{1}{m(m-1)}\sum_{i=1}^{m}\sum_{j=1,j\ne i}^{m}k(X_{i},X_{j}) + \frac{1}{n(n-1)}\sum_{i=1}^{n}\sum_{j=1,j\ne i}^{n}k(Y_{i},Y_{j}) \notag \\
&\ \ \ \ \ - \frac{2}{mn}\sum_{i=1}^{m}\sum_{j=1}^{n}k(X_{i},Y_{j}), \\
\textrm{MMD}^2_{b} &= \frac{1}{m^2}\sum_{i=1}^{m}\sum_{j=1}^{m}k(X_{i},X_{j}) + \frac{1}{n^2}\sum_{i=1}^{n}\sum_{j=1}^{n}k(Y_{i},Y_{j}) - \frac{2}{mn}\sum_{i=1}^{m}\sum_{j=1}^{n}k(X_{i},Y_{j}).
\end{align}
Here, $\textrm{MMD}^2_{u}$ is an unbiased estimator of $\textrm{MMD}^2(P,Q)$ and is in general preferred over $\textrm{MMD}^2_{b}$. When the kernel $k$ is characteristic, such as the Gaussian kernel or the Laplacian kernel, the MMD behaves as a metric \citep{sriperumbudur2010hilbert}.

\cite{gretton2007kernel} studied the asymptotic properties of $\textrm{MMD}^2_{u}$ and found that $\textrm{MMD}^2_{u}$ degenerated under the null hypothesis of equal distributions. They then considered $m\textrm{MMD}^2_{u}$ when $m=n$ and showed that $m\textrm{MMD}^2_{u}$ converges to $\sum_{l=1}^{\infty}\lambda_{l}(z_{l}^2-2)$ under $H_{0}$. Here $z_{l}\stackrel{iid}{\sim} N(0,2)$ and the $\lambda_{l}$s are solutions of the eigenvalue equation
$\int_{\mathcal{X}}k_c(X,X')\psi_{l}(X)dP(X) = \lambda_{l}\psi_{l}(X')$
with $k_c(X_{i},X_{j}) = k(X_{i},X_{j})-E_{X}k(X_{i},X)-E_{X}k(X,X_{j})+E_{X,X'}k(X,X')$, the centred RKHS kernel. Since the limiting distribution $\sum_{l=1}^{\infty}\lambda_{l}(z_{l}^2-2)$ is an infinite sum, a few approaches were proposed to approximate it: a moment matching approach using Pearson curves \citep{gretton2007kernel}, a spectrum approximation approach, and a Gamma approximation approach \citep{gretton2009fast}. However, these approaches have their limitations. For example, \cite{gretton2009fast} noted that the performance of tests based on the moment matching  and  Gamma approximation methods is not guaranteed. In addition, all these approaches only apply to balanced sample designs, where the sample sizes of the two samples are equal. Consequently, in terms of guaranteed performance of the test and for possibly unbalanced sample sizes, a commonly preferred approach in many applications is the use of bootstrapping to approximate the $p$-value, despite its high computational cost \citep{gretton2009fast, li2019optimality}. 

\cite{gretton2012optimal} studied the choice of the kernel and  bandwidth parameter to maximize the power of the test from a set of linear combinations of Gaussian kernels in a training set. More recently, \cite{ramdas2015adaptivity} found that the power of the test based on the Gaussian kernel remains independent of the kernel bandwidth when the bandwidth surpasses the median of all pairwise distances among observations. Therefore, in the following, without further specification, we employ the most popular characteristic kernel, the Gaussian radial basis function kernel, with the median heuristic as the bandwidth parameter. 


\subsection{A problem of $\textrm{MMD}^2_{u}$} \label{sec:prob:lim}

Even though $\textrm{MMD}^2_{u}$ works well under many settings, it exhibits some weird behaviors under some common alternatives. Consider a simple example involving Gaussian data: $X_{1},\ldotp\ldotp\ldotp,X_{50} \stackrel{iid}{\sim} N_{d}(0_{d},\Sigma(0.4))$; $Y_{1},\ldotp\ldotp\ldotp,Y_{50} \stackrel{iid}{\sim} N_{d}(a1_{d},b\Sigma(0.4))$, where $\Sigma(0.4)$ is a $d\times d$ matrix with its $(i,j)$-th element $0.4^{|i-j|}$, $0_{d}$ is a $d$-dimensional vector of zeros,  $1_{d}$ is a vector of ones, and $d=50$. We explore three settings:
\begin{itemize}
	\item Setting 1: $a = 0.21, b = 1$.
	\item Setting 2: $a = 0.21, b = 1.04$.
	\item Setting 3: $a = 0, b = 1.1$.
\end{itemize}
Table \ref{table:poor} presents the estimated power of the $\textrm{MMD}^2_{u}$ test based on 1,000 simulation runs. In each simulation run, 10,000 bootstrap replicates are used to approximate the $p$-value. We refer to this test `MMD-Bootstrap' for simplicity.  It is notable that MMD-Bootstrap performs admirably in detecting mean differences in Setting 1.  However, in Setting 2, despite the presence of an additional variance difference, it exhibits slightly lower power than in Setting 1.  In Setting 3, where the difference pertains only to the variance, MMD-Bootstrap performs poorly.
\begin{table}[h]
	\caption{\label{table:poor}Estimated power (by 1,000 trials) of MMD-Bootstrap at 0.05 significance level}
	\centering
	\begin{tabular}{ccc}
		Setting 1 & Setting 2 & Setting 3\\
		0.912     & 0.886 & 0.071    \\
	\end{tabular}
\end{table}

\begin{figure}[h]
	\centering
	\includegraphics[width=0.95\columnwidth]{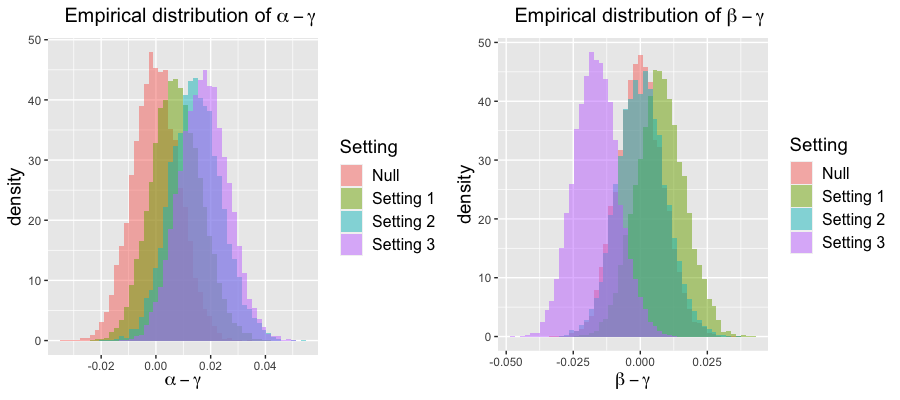}
	\caption{Empirical distributions of $\alpha-\gamma$ and $\beta-\gamma$ based on 10,000 simulation runs under settings 1, 2, 3 and the null of no distribution difference $(a=0, b=1)$.}
	\label{fig:empirical}
\end{figure}

To delve into the underlying causes of these observations, we examine the empirical distributions of $\alpha-\gamma$ and $\beta-\gamma$, where $\alpha = (m^2-m)^{-1}\sum_{i=1}^{m}\sum_{j=1,j\ne i}^{m}k(X_{i},X_{j})$, $\beta = (n^2-n)^{-1}\sum_{i=1}^{n}\sum_{j=1,j\ne i}^{n}k(Y_{i},Y_{j})$, $\gamma = (mn)^{-1}\sum_{i=1}^{m}\sum_{j=1}^{n}k(X_{i},Y_{j})$ ($\textrm{MMD}^2_{u} = \alpha + \beta - 2\gamma$).  Figure \ref{fig:empirical} provides a visual representation of our findings.  In Setting 1,  the distributions of $\alpha-\gamma$ and $\beta-\gamma$ exhibit a noticeable rightward shift compared to those under the null hypothesis.  Consequently, $\textrm{MMD}^2_{u}$ tends to yield large values in Setting 1, resulting in a test power of  0.912. In Setting 2, when an  additional variance is introduced, the empirical distribution of $\alpha-\gamma$ indeed shifts further to the right. However, the empirical distribution of $\beta-\gamma$ remains similar to that under the null. Consequently, in Setting 2, the effects of $\alpha-\gamma$ and $\beta-\gamma$ offset each other, leading to a lower test power compared to Setting 1. This phenomenon becomes more pronounced in Setting 3, where $\beta-\gamma$ is predominantly negative and almost entirely cancels out $\alpha-\gamma$, resulting in a test power of merely 0.071. 

On the other hand,  Figure \ref{fig:empirical} clearly illustrates that in Setting 3, $\alpha-\gamma$ and $\beta-\gamma$ do deviate from the null hypothesis (depicted in purple versus pink). The magnitude of these derivations in Setting 3 surpasses that in Setting 1 for both $\alpha-\gamma$ and $\beta-\gamma$.  However, it is crucial to note that these deviations occur in opposite directions, causing the test statistic $\textrm{MMD}^2_{u}$ to struggle to capture the underlying signal.



\subsection{Related works and our contribution}

There have been several attempts in two-sample testing problems based on the MMD. \cite{balasubramanian2021optimality} contended that potent tests could be devised through regularized kernel embeddings.  However, this method relies on  knowledge of distribution $P$, making it impractical for general two-sample testing problems. In response to this limitation, \cite{li2019optimality} showcased that a test employing the Gaussian kernel with a properly chosen scaling parameter achieves minimax optimality against smooth alternatives. \cite{chakraborty2021new} proposed a new metric designed for  low-dimensional settings and extended its application to  high-dimensional scenarios. \cite{gao2021two} proposed a studentized sample MMD and studied the asymptotic theory as both the sample size and dimension approach infinity. We evaluated these methods in the three settings in Section \ref{sec:prob:lim} and the results are summarized in Table \ref{table:setting_new}: AG (Adaptive Gaussian kernel method, \cite{li2019optimality}), $\textrm{ND}_{1}$ and $\textrm{ND}_{2}$ (New Distance, \cite{chakraborty2021new}), where 1 and 2 represent different choices of a key parameter used in their numerical studies, and TM (Studentized MMD, \cite{gao2021two}). 
\begin{table}[h]
	\caption{\label{table:setting_new}Estimated power of the tests at 0.05 significance level}
	\centering
	\begin{tabular}{cccc}
		& Setting 1 & Setting 2 & Setting 3\\
        AG & 0.093 & 0.133 & 0.091 \\
		$\textrm{ND}_{1}$ & 0.939 & 0.912 & 0.076    \\
		$\textrm{ND}_{2}$ & 0.846 & 0.832 & 0.081    \\
	    TM & 0.868 & 0.852 & 0.087    \\
	\end{tabular}
\end{table}

We observe that $\textrm{ND}_{1}$, $\textrm{ND}_{2}$, and TM exhibit a similar behavior to the MMD, displaying slightly lower power in Setting 2 compared to Setting 1 and poor performance in Setting 3.  Conversely, AG does not perform effectively in these settings.

In response to the findings in Section \ref{sec:prob:lim}, we take a novel approach.  We delve deeper into the behavior of $\alpha$ and $\beta$ under the permutation null distribution and propose a new statistic, denoted as GPK, that considers deviations in both directions, aiming to accommodate a broader spectrum of high-dimensional alternatives compared to $\textrm{MMD}^2_{u}$.  Additionally, we introduce a test statistic, fGPK, which operates similar to GPK but with fast type I error control.  Utilizing a similar methodology, we further develop $\textrm{fGPK}_{\textrm{M}}$ that has power on par and sometimes much better than prevailing MMD-based tests and at the same time with fast type I error control.  It is noteworthy that  these newly proposed tests, GPK, fGPK, and $\textrm{fGPK}_{\textrm{M}}$, work for both equal and unequal sample sizes. The new methods are implemented in an $\texttt{R}$ package $\texttt{kerTests}$, and the codes for reproducing simulation results can be found at \url{https://github.com/hoseungs/kerTests}.


\section{A New Test Statistic} \label{sec:new}

\subsection{Notation}

Let $N=m+n$ be the total sample size.  We work under the permutation null distribution that assigns a probability of $m!n!/N!$  to each of the $N!/m!/n!$ permutations of the sample labels.  In the following, $\pr$, $\ep$, $\var$, and $\cov$ represent the probability,  expectation,  variance under the permutation null distribution.  Conversely, we use $E$ and $Pr$ to denote the expectation and probability, respectively, with respect to the probability distribution.


We denote the combined observations from the two samples as $Z_{1},\ldots,Z_{N}$. Let $k_{ij} = k(Z_{i},Z_{j})$ for $i,j = 1, \ldots, N$,  $\bar{k} = \sum_{i=1}^{N}\sum_{j=1,j\ne i}^{N}k_{ij}/(N^2-N)$, $\tilde{k}_{ij} = (k_{ij} - \bar{k})I_{i\ne j}$, and $\tilde{k}_{i\cdot} = \sum_{j=1,j\ne i}^{N}\tilde{k}_{ij}$ for $i=1,\ldots,N$. We write $a_N = O(b_N)$ when $a_N$ has the same order as $b_N$, and $a_N = o(b_N)$ when $a_N$ is dominated by $b_N$ asymptotically, i.e., $\lim_{N\rightarrow\infty}(a_{N}/b_{N}) = 0$.


\subsection{A pattern in moderate/high dimensions} \label{sec:motivation}

\begin{figure}[h]
	\centering
	\includegraphics[width=0.95\columnwidth]{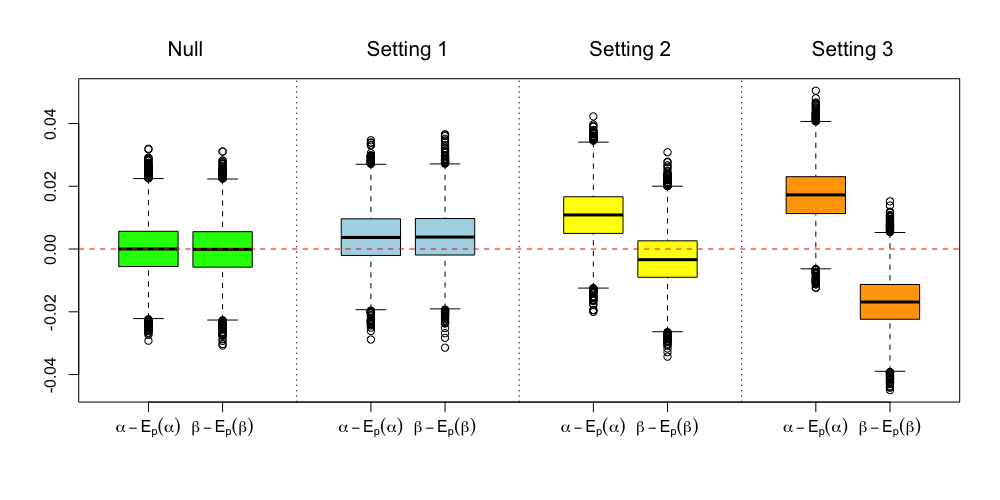}
	\vskip -0.3in
	\caption{Boxplots of $\alpha-\ep(\alpha)$ and $\beta-\ep(\beta)$ of 10,000 simulated datasets under null $(a=0, b=1)$, Setting 1 $(a=0.21, b=1)$, Setting 2 $(a=0.21, b=1.04)$, and Setting 3 $(a=0, b=1.1)$.}
	\label{fig:heat}
\end{figure}

Figure \ref{fig:heat} presents boxplots of $\alpha - \ep(\alpha)$ and $\beta-\ep(\beta)$ from 10,000 simulated datasets under the three settings in Section \ref{sec:prob:lim}, as well as under the null hypothesis $(a=0, b=1)$. In Setting 1, we see that both $\alpha$ and $\beta$ tend to be larger than their null expectations, which is consistent with $\textrm{MMD}^2_{u}$ being large. In Setting 2, $\alpha$ still tends to be larger than its null expectation, while $\beta$ tends to be smaller than its null expectation, which could cause the effect of $\alpha$ and $\beta$ in $\textrm{MMD}^2_{u}$ to offset each other. This phenomenon gets severer in Setting 3. The reason this happens lies in the curse of dimensionality, where the volume of a $d$-dimensional space experiences exponential growth with increasing $d$.  Consequently, observations from a distribution with larger variance can become widely dispersed and tend to find themselves closer to observations from the distribution with smaller variance.  This can lead to either $\alpha$ or $\beta$ being smaller than its null expectation, depending on which sample has larger variance.


\subsection{A generalized permutation-based kernel two-sample test statistic} \label{sec:newstat}

Based on the findings in Section \ref{sec:motivation}, we segregate $\alpha$ and $\beta$ and propose the following statistic:
\begin{equation}
\textrm{GPK} = \big(\alpha - \ep(\alpha), \beta-\ep(\beta)\big)\Sigma_{\alpha,\beta}^{-1}\left(\begin{tabular}{c}
$\alpha - \ep(\alpha)$ \\
$\beta-\ep(\beta)$
\end{tabular}\right),
\end{equation}
where $\Sigma_{\alpha,\beta} = \var((\alpha,\beta)^{T})$. The expressions of $\ep(\beta)$, $\ep(\beta)$, and $\Sigma_{\alpha,\beta}$ can be derived analytically and they are provided in Theorem \ref{thm:mean}. The new test statistic designed in this way aggregates deviations of $\alpha$ and $\beta$ from their expectations under the permutation null in both directions, so it can cover more general alternatives than $\textrm{MMD}^2_{u}$.

\begin{figure}[h]
	\centering
	\includegraphics[width=\columnwidth]{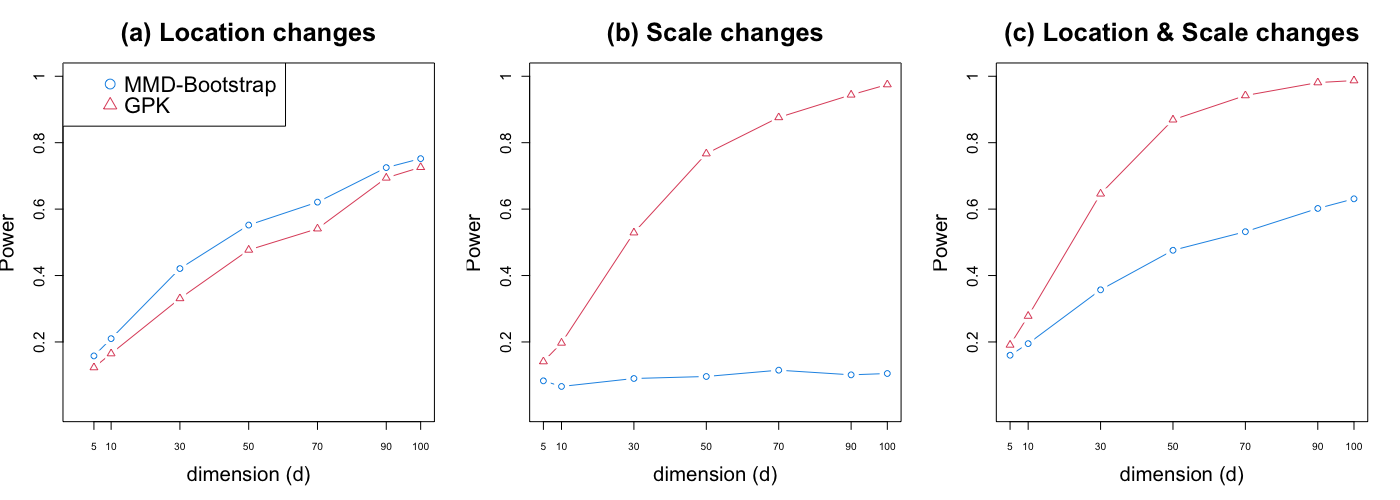}
	\caption{Estimated power of MMD-Bootstrap (o) and GPK ($\triangle$) at 0.05 significance level for multivariate Gaussian data: (a) $a=0.15$, $b=1$, (b) $a=0$, $b=1.15$, (c) $a=0.12$, $b=1.15$.}
	\label{fig:compare}
\end{figure}

To assess the performance of GPK, we conduct preliminary evaluations (further simulation studies are detailed in Section \ref{sec:sim}).  We adopt a simulation setting akin to that in Section \ref{sec:prob:lim}, involving Gaussian data $N_{d}(0_{d},\Sigma(0.4))$ vs. $N_{d}(a1_{d},b\Sigma(0.4))$ with $m=n=50$, and considering location and$\slash$or scale alternatives. The estimated power of GPK is compared to that of MMD-Bootstrap across 1,000 trials, and the results are presented in Figure \ref{fig:compare}.  We see that GPK has comparable power to MMD-Bootstrap for location alternatives.  However, when the difference involves scale, MMD-Bootstrap performs poorly, while GPK demonstrates significantly higher power.  When both mean and variance differ, GPK in general outperforms MMD-Bootstrap.

The analytic formulas for $\ep(\alpha)$, $\ep(\beta)$, and $\Sigma_{\alpha,\beta}(i,j)$, the $(i,j)$-th element of $\Sigma_{\alpha, \beta}$, are provided in the following theorem.
\begin{theorem} \label{thm:mean}
	Under the permutation null distribution, we have
	\begin{flalign*}
	&\ep(\alpha) = \ep(\beta) =  \bar{k},   \\
	&\Sigma_{\alpha, \beta}(1,1) = \left\{2A f_1(m)+4B f_2(m)+C f_3(m)\right\}/m^2/(m-1)^2 - \bar{k}^2, \\
	&\Sigma_{\alpha, \beta}(2,2) = \left\{2A f_1(n) +4B f_2(n) +C f_3(n)\right\}/n^2/(n-1)^2 - \bar{k}^2, \\
	&\Sigma_{\alpha, \beta}(1,2) = \Sigma_{\alpha, \beta}(2,1) = C\{N(N-1)(N-2)(N-3)\}^{-1} - \bar{k}^2, 
	\end{flalign*}
	where
	\begin{align*}
	& f_1(x)  = \frac{x(x-1)}{N(N-1)}, \ \ f_2(x)  = \frac{x(x-1)(x-2)}{N(N-1)(N-2)}, \ \ f_3(x) = \frac{x(x-1)(x-2)(x-3)}{N(N-1)(N-2)(N-3)}, \\
	& A = \sum_{i=1}^{N}\sum_{j=1,j\ne i}^{N}k^2_{ij}, \ \ 
	B = \sum_{i=1}^{N}\sum_{j=1,j\ne i}^{N}\sum_{u=1,u\ne j, u\ne i}^{N}k_{ij}k_{iu}, \\ 
	& C = \sum_{i=1}^{N}\sum_{j=1,j\ne i}^{N}\sum_{u=1,u\ne j, u\ne i}^{N}\sum_{v=1,v\ne u, v\ne j, v\ne i}^{N}k_{ij}k_{uv}.
	\end{align*}
\end{theorem}
To prove this theorem, we rewrite $\alpha$ and $\beta$ in the following way. For each  $Z_{1},\ldotp\ldotp\ldotp,Z_{N}$, let $g_{i} = 0$ if observation $Z_{i}$ is from sample $X$ and $g_{i} = 1$ if observation $Z_{i}$ is from sample $Y$. Then, 
\begin{align}
\alpha &= \frac{1}{m(m-1)}\sum_{i=1}^{m}\sum_{j=1,j\ne i}^{m}k(X_{i},X_{j}) =  \frac{1}{m(m-1)}\sum_{i=1}^{N}\sum_{j=1,j\ne i}^{N}k_{ij}I_{g_{i}=g_{j}=0},\\
\beta &= \frac{1}{n(n-1)}\sum_{i=1}^{n}\sum_{j=1,j\ne i}^{n}k(Y_{i},Y_{j}) = \frac{1}{n(n-1)}\sum_{i=1}^{N}\sum_{j=1,j\ne i}^{N}k_{ij}I_{g_{i}=g_{j}=1}.
\end{align}
Therefore, the computation of $\ep(\alpha)$ essentially reduces to computing $\ep(I_{g_{i}=g_{j}=0})$, and similar for $\ep(\beta)$. Calculating the variance of $\alpha$ and the covariance of $\alpha$ and $\beta$ under the permutation null distribution necessitates a more meticulous examination of various combinations. The detailed proof of the theorem is provided in Supplement A.

\begin{theorem} \label{thm:det}
	For $m,n\ge2$, the proposed statistic GPK is well-defined except in cases where the $k_{ij}$ values fall into either of the following two corner cases:
	\begin{enumerate}[label=(C\arabic*)]
		\item All $\sum_{j=1,j\ne i}^{N}k_{ij}$ values are identical for $i=1,\ldotp\ldotp\ldotp,N$. \label{condition1}
		\item All $\sum_{j=1,j\ne i}^{N}k_{ij} - (N-2)k_{iN}$ values are identical for $i=1,\ldotp\ldotp\ldotp,N-1$. \label{condition2}
	\end{enumerate}
\end{theorem}

The proof of Theorem \ref{thm:det} can be established through mathematical induction, with the details  in Supplement B. It is difficult to simplify (C1) and (C2) further, while the two corner cases are rare to happen.  We illustrate this rarity through simulations, as outlined in Supplement C.


\section{Asymptotics and Alternative Tests} \label{sec:newtest}


\subsection{A decomposition of GPK and asymptotic results} \label{sec:asymp}

Given the new test statistic GPK, the next question is to compute the $p$-value of the test. Although permutation of sample indices is a viable option, it can be time-consuming.  Therefore, we aim to investigate the asymptotic distribution of GPK under the permutation null distribution.  We observe that GPK can be decomposed to the squares of two uncorrelated quantities.   One of these quantities converges to a Gaussian distribution under certain mild conditions, while  the other is closely tied to $\textrm{MMD}^2_{u}$.  Furthermore, a modified form of the quantity  related to $\textrm{MMD}^2_{u}$ also converges to a Gaussian distribution under certain mild conditions. Based on these findings, we propose two tests, fGPK and $\textrm{fGPK}_{\textrm{M}}$, for which we can approximate $p$-values using analytic formulas.  The former is closely linked to the test based on GPK, while the latter is related to the test based on $\textrm{MMD}^2_{u}$.





\begin{theorem} \label{thm:decomp}
	The statistic GPK can be decomposed as follows:
	$$\textrm{GPK} = Z_{W}^2 + Z_{D}^2,$$
	where 
	$$	Z_{W} = \frac{W-\ep(W)}{\surd \var(W)}, \
	Z_{D} = \frac{D-\ep(D)}{\surd \var(D)},$$
	with $W = m\alpha/N + n\beta/N$ and $D = m(m-1)\alpha/N/(N-1) - n(n-1)\beta/N/(N-1)$. 
\end{theorem}
The complete proof for this theorem is provided in Supplement D.

\begin{remark}\label{remark:WD}
	The analytic expressions for the expectation and variance of $W$ and $D$ can be readily obtained from Theorem \ref{thm:mean}:
	\begin{align*}
	\ep(W) & = \bar{k}, \ \ \ep(D) = (m-n)\bar{k}/N, \\
	\var(W) & = \frac{mn\left\{(N-2)2A+2(2A+4B+C)/(N-1)-(4A+4B)\right\}}{N^3(N-1)(N-3)(m-1)(n-1)}, \\
	\var(D) & = \frac{mn(N-4)\left\{(4A+4B)-4(2A+4B+C)/N\right\}}{N^3(N-1)^3(N-3)}.
	\end{align*}
\end{remark}

\begin{remark} \label{remark:permu}
	The quantity $Z_{W}$ is closely linked to $\textrm{MMD}^2_{u}$. Given that
	$m(m-1)\alpha + n(n-1)\beta + 2mn\gamma =\sum_{i=1}^{N}\sum_{j=1,j\ne i}^{N}k_{ij} = N(N-1)\bar{k}$, we have
	\begin{align*}
	\textrm{MMD}_{u}^2 &= \alpha + \beta - 2\gamma = \alpha + \beta - (mn)^{-1}\left\{N(N-1)\bar{k} - m(m-1)\alpha - n(n-1)\beta\right\} \\
	& = (mn)^{-1}N(N-1)(W-\bar{k}) = (mn)^{-1}N(N-1)(W-\ep(W)).
	\end{align*}
	Thus, the test statistic $Z_{W}$ is essentially equivalent to MMD-Permutation -- the $\textrm{MMD}_{u}^2$ test with its $p$-value computed under the permutation null distribution.  So GPK could in general deal with the alternatives that $\textrm{MMD}^2_{u}$ covers. In addition, $Z_{D}$ covers a new region of alternatives that might not be captured by $\textrm{MMD}^2_{u}$, making GPK work for more general alternatives.  
\end{remark}

We now investigate the consistency of GPK.
\begin{theorem} \label{thm:consistency}
    When the Gaussian kernel is used, as $N\rightarrow\infty$, $m/N \rightarrow p \in (0,1)$, $W-\ep\left(W\right)$ converges in probability to 0 when $P=Q$ and to a positive constant when $P\ne Q$, and $D-\ep\left(D\right)$ converges in probability to $\eta: = 2mnN^{-2}<(m-1)\mu_{P}/(N-1)+(n-1)\mu_{Q}/(N-1), \mu_{P}-\mu_{Q}>_{H}$, where $\mu_{P}$ and $\mu_{Q}$ are mean embeddings of $P$ and $Q$ in the RKHS, respectively.
\end{theorem}
The proof is provided in Supplement E. The $D-\ep\left(D\right)$ part is not as straightforward as the $W-\ep\left(W\right)$ part. When $P=Q$, it is easy to see that $\eta=0$. When $P\ne Q$, $\eta$ is a non-zero constant for most of the time. Nevertheless, the consistency of GPK $(= Z_{D}^2+Z_{W}^2)$ can still be ensured based on Theorem \ref{thm:consistency}.

We next examine the asymptotic permutation null distribution of the statistics.  The limiting distribution of $m\textrm{MMD}_{u}^2$ is challenging to handle \citep{gretton2007kernel}.  Due to the intrinsic relation between $\textrm{MMD}_{u}^2$ and $W$, it is also difficult to handle the limiting distribution of $W$.  Hence, we work on a related quantity. Let $W_{r} = rm\alpha/N + n\beta/N$ be a weighted version of $W$, where $r$ is a constant. Note that $W_{1} = W$.  Similar to $Z_W$, we define
$$Z_{W,r} = \frac{W_{r}-\ep(W_{r})}{\surd{\var(W_{r})}}. $$


We work under the following two conditions.
\begin{condition}\label{cond1}
	$\sum_{i=1}^{N}|\tilde{k}_{i\cdot}|^s = o\left(\{\sum_{i=1}^{N}\tilde{k}_{i\cdot}^2\}^{s/2}\right)$ for all integers $s>2$.
\end{condition}
\begin{condition}\label{cond2}
	$\sum_{i=1}^{N}\sum_{j=1,j\ne i}^{N}\tilde{k}_{ij}^2 = o\left(\sum_{i=1}^{N}\tilde{k}_{i\cdot}^2\right)$.
\end{condition}

\begin{theorem} \label{asymp:zscal}
	Under the permutation null distribution, as $N\rightarrow\infty$, $m/N \rightarrow p \in (0,1)$, $Z_{D}$ converges in distribution to the standard normal distribution under Condition \ref{cond1}, and $Z_{W,r}$ converges in distribution to the standard normal distribution under Conditions \ref{cond1} and \ref{cond2} when $r\ne 1$.
\end{theorem}
The proof for this theorem can be found in Supplement F.  


\begin{remark}
	Condition \ref{cond1} can be satisfied when $|\tilde{k}_{i\cdot}| = O(N^{\delta})$ for a constant $\delta$, $\forall i$, and Condition \ref{cond2} would further be satisfied if we also have  $\tilde{k}_{ij} = O(N^\kappa)$ for a constant $\kappa<\delta-0.5, \forall i,j$. Noting that a special case in which both conditions are satisfied is when the $\tilde{k}_{ij}$ values are of constant order.  This is typically the case when using the Gaussian kernel with the median heuristic, unless there are exceptional circumstances involving significant outliers. 
\end{remark}

\begin{figure}[h]
	\centering
	\includegraphics[width=\columnwidth]{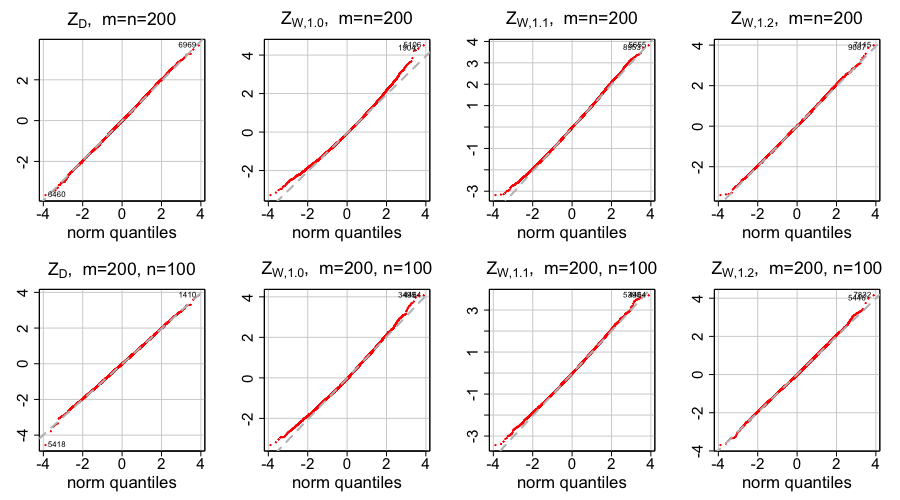}
	\caption{Normal quantile-quantile plots (red dots) of $Z_{D}$, $Z_{W,1.0}$, $Z_{W,1.1}$, $Z_{W,1.2}$ with the gray dashed line the baseline goes through the origin and of slope 1.}
	\label{qqplot}
\end{figure}

Figure \ref{qqplot} shows normal quantile-quantile plots for $Z_{D}$, $Z_{W, 1.0}$, $Z_{W, 1.1}$, and $Z_{W, 1.2}$ from 10,000 permutations under different choices of $m$ and $n$ for Gaussian data with $d=100$. We see that, when $m$,$n$ are in the hundreds, the permutation distributions can already be well approximated by the standard normal distribution for $Z_{D}$ and for $Z_{W,r}$ with $r$ away from 1, such as $r=1.2$. 

\subsection{Fast tests: fGPK and fGPK$_M$} \label{sec:fasttest}

While $Z_{W,r}$, $r\ne 1$, converges to the standard normal distribution under mild conditions, its performance deteriorates as $r$ deviates from 1 under the location alternative. Table \ref{table:simes} displays the estimated power (based on 100 simulation runs) of $Z_{W,r}$ for Gaussian data $N_{d}(\mu_{1}, I_{d})$ vs. $N_{d}(\mu_{2}, I_{d})$.
The $p$-value of each test is approximated using 10,000 permutations for a fair comparison.  The results reveal a decrease in test power as $r$ moves away from 1.  To harness the asymptotic findings and maximize power, we propose employing a Bonferroni test on $Z_{W,1.2}$, $Z_{W,0.8}$, and $Z_{D}$. We select $Z_{W,1.2}$ and $Z_{W,0.8}$ because they are reasonably Gaussian distributed under finite sample sizes, while maintaining relatively good power (in terms of location alternatives).  Let $p_{W,1.2}$, $p_{W,0.8}$, and $p_{D}$ denote the approximated $p$-value of the test that rejects for large values of $Z_{W,1.2}$, $Z_{W,0.8}$, and $|Z_{D}|$, respectively, based on their limiting distributions, i.e., if the values of $Z_{W,1.2}$, $Z_{W,0.8}$, and $Z_{D}$ are $b_{W,1.2}$, $b_{W,0.8}$, and $b_{D}$, respectively, then $p_{W,1.2} = 1 - \Phi(b_{w,1.2})$, $p_{W,0.8}= 1 - \Phi(b_{w,0.8})$, and $p_{D} = 2\Phi(-|b_{D}|)$. Then, fGPK rejects the null hypothesis if $3\min(p_{D}, p_{W,1.2}, p_{W,0.8})$ is less than the significance level. 

\begin{table}
	\caption{\label{table:simes}Estimated power of $Z_{W,r}$ at 0.05 significance level, $m=n=100$, $\Delta = \|\mu_{1}-\mu_{2}\|_{2}$}
	\centering
	\begin{tabular}{ccccccc}
		$d$ & 10 & 30 & 50 & 70 & 90 & 100  \\
		$\Delta$ & 0.3 & 0.5 & 0.7 & 0.8 & 0.9 & 1.0 \\
		$r=1.3$ & 0.11 & 0.24 & 0.36 & 0.36 & 0.49 & 0.50   \\
		$r=1.2$ & 0.15 & 0.28 & 0.43 & 0.50 & 0.68 & 0.63   \\
		$r=1.1$ & 0.10 & 0.42 & 0.55 & 0.70 & 0.83 & 0.84   \\
		$r=1.0$ & 0.25 & 0.52 & 0.60 & 0.77 & 0.90 & 0.86   \\
		$r=0.9$ & 0.22 & 0.47 & 0.41 & 0.77 & 0.76 & 0.78   \\
		$r=0.8$ & 0.16 & 0.36 & 0.27 & 0.49 & 0.57 & 0.54   \\
		$r=0.7$ & 0.15 & 0.23 & 0.20 & 0.37 & 0.32 & 0.33   \\
	\end{tabular}
\end{table}





Similarly, $\textrm{fGPK}_{\textrm{M}}$ is defined to reject the null hypothesis if $2\min(p_{W,1.2}, p_{W,0.8})$ is less than the significance level to approximate the MMD-permutation test. We expect $\textrm{fGPK}_{\textrm{M}}$ to be powerful for location alternatives. 

Next we examine the consistency of the test based on $W_{r}$.
\begin{theorem}
    When the Gaussian kernel is used, as $N\rightarrow\infty$, $m/N \rightarrow p \in (0,1)$, $W_{0.8}-\ep\left(W_{0.8}\right)$ and $W_{1.2}-\ep\left(W_{1.2}\right)$ converge in probability to 0 when $P=Q$ and and at least one of them converges to a positive constant when $P\ne Q$.
\end{theorem}
The proof is provided in Supplement G. 

We compare the computational cost of the two fast tests, fGPK and fGPK$_M$, with MMD-Pearson and MMD-Bootstrap.  Notice that MMD-Pearson can only be applied to equal sample sizes, so we set $m=n$.  Both samples are drawn from the standard 100-dimensional Gaussian distribution. Table \ref{tab:runtime:matlab} reports the time cost of the methods implemented in $\texttt{Matlab}$. For MMD-Pearson and MMD-Bootstrap, we use the $\texttt{Matlab}$ codes released by Arthur Gretton, publicly available at \url{http://www.gatsby.ucl.ac.uk/~gretton/mmd/mmd.htm}. The time comparison for these methods implemented in $\texttt{R}$ can be found in Supplement H. It is not surprising to see that $\textrm{fGPK}_{\textrm{M}}$ and fGPK are much faster than MMD-Bootstrap, while they are also much faster than MMD-Pearson, especially when the sample size is large. 
\begin{table}[htp!]
	\caption{\label{tab:runtime:matlab}Average computation time in seconds (standard deviation) from 10 simulation runs for each $m$. All experiments were conducted using $\texttt{Matlab}$ on a 2.2 GHz Intel Core i7 processor.}
	\centering
	\begin{tabular}{ccccc}
		$m$ & 100 & 250 & 500 & 1000 \\
		$\textrm{fGPK}_{\textrm{M}}$ & 0.001 (0.000) &  0.005 (0.001) & 0.021 (0.001) &  0.105 (0.004)  \\
		$\textrm{fGPK}$ & 0.002 (0.002) &  0.004 (0.000) & 0.022 (0.001) &  0.105 (0.003)  \\
		MMD-Pearson  &   0.012 (0.010) &   0.093 (0.002) & 0.739 (0.037) & 13.13 (0.88) \\
		MMD-Bootstrap & 1.477 (0.048) & 8.168 (0.177) & 37.44 (5.13) & 251.9 (16.1) \\
	\end{tabular}
	
\end{table}


\section{Simulation Studies} \label{sec:sim}

In this section, we compare the three new tests (GPK, fGPK, $\textrm{fGPK}_{\textrm{M}}$) with two commonly used MMD-based tests (MMD-Pearson and MMD-Bootstrap) on a variety of settings in moderate/high dimensions. We also include other nonparametric tests that utilize the ball divergence (BT) \citep{pan2018ball}, classifier (CT) \citep{lopez2016revisiting}, and graphs (GT)  \citep{chen2017new},  implemented through $\texttt{R}$ packages $\texttt{ball}$, $\texttt{Ecume}$, and $\texttt{gTests}$, respectively. For the GT test, we employ the 5-MST (minimum spanning tree). We consider the following settings:
\begin{itemize}
	\item Multivariate Gaussian data: $N_{d}(0_{d},\Sigma(0.4))$ vs.  $N_{d}(a1_{d},\sigma^2\Sigma(0.4))$. 
	\item Multivariate $t$-distributed data: $t_{20}(0_{d},\Sigma(0.4))$ vs.  $t_{20}(a1_{d},\sigma^2\Sigma(0.4))$. 
	\item Chi-square data: $(\Sigma(0.4))^{1/2}u_1$ vs.  $(\sigma^2\Sigma(0.4))^{1/2}u_2+ a1_{d}$, 
	where $u_1$ and $u_2$ are length-$d$ vectors with each component i.i.d. from the $\chi_{3}^2$ distribution. 
\end{itemize}
For multivariate Gaussian data, we also compare the tests under unbalanced settings $(m\ne n)$.  Sparse mean and variance change scenarios are considered as well, with additional details provided in Supplement I. 
For each simulation setting, we consider various dimensions.  We select distribution parameters such that the tests exhibit moderate power for comparison purposes. The significance level is set to be 0.05 for all tests. The estimated power (by 1,000 simulation runs) is presented in Tables \ref{tab:loc,scal1} -- \ref{tab:chi}.  In these tables,  $\Delta = \|a1_{d}\|_{2}$.

\begin{table}[h]
	\caption{\label{tab:loc,scal1}Estimated power of the tests for multivariate Gaussian data $(m=n=50)$}
	\centering
	\begin{tabular}{ccccc}
		& \multicolumn{4}{c}{Location Alternatives ($\Delta$)}  \\
		$d$ & 50 & 100 & 500 & 1000  \\
		$\Delta$ $|$ $\sigma^2$ & 1.13 & 1.50 & 2.23 & 2.84 \\
		MMD-Pearson & 0.177 & 0.155 & 0.006 & 0.002   \\
		MMD-Bootstrap &  0.651 & 0.801 & 0.516 & 0.334   \\
		GPK &  0.567 & 0.761 & 0.772 & 0.891   \\
		fGPK &  0.527 & 0.704 & 0.747 & 0.868   \\
		$\textrm{fGPK}_{\textrm{M}}$ & 0.578 & 0.749 & 0.800 & 0.905   \\
		BT &  0.362 & 0.384 & 0.216 & 0.222   \\
		CT &  0.367 & 0.464 & 0.525 & 0.635   \\
		GT &  0.193 & 0.282 & 0.303 & 0.388   \\
	\end{tabular} 
	\centering
	\begin{tabular}{cccc}
		\multicolumn{4}{c}{Scale Alternatives ($\sigma^2$)}  \\
		50 & 100 & 500 & 1000  \\
		1.11 & 1.09 & 1.05 & 1.04 \\
		0.001 & 0.001 & 0.000 & 0.000  \\
		0.065 & 0.042 & 0.001 & 0.000   \\
		0.472 & 0.611 & 0.843 & 0.913   \\
		0.460 & 0.605 & 0.848 & 0.900  \\
		0.317 & 0.432 & 0.612 & 0.702   \\
		0.534 & 0.686 & 0.890 & 0.941  \\
		0.074 & 0.040 & 0.023 & 0.018   \\
		0.370 & 0.418 & 0.659 & 0.706   \\
	\end{tabular}
\end{table}
\begin{table}[h]
	\caption{\label{tab:loc,scal2}Estimated power of the tests for multivariate Gaussian data $(m=100, n=50)$}
	\centering 	
	\begin{tabular}{ccccc}
		& \multicolumn{4}{c}{Location Alternatives ($\Delta$)}  \\
		$d$ & 50 & 100 & 500 & 1000  \\
		$\Delta$ $|$ $\sigma^2$ & 0.98 & 1.30 & 2.01 & 2.84 \\
		MMD-Pearson &  - &- & -& -  \\
		MMD-Bootstrap &  0.612 & 0.632 & 0.132 & 0.085   \\
		GPK &  0.620 & 0.733 & 0.817 & 0.979   \\
		fGPK &  0.529 & 0.673 & 0.770 & 0.964   \\
		$\textrm{fGPK}_{\textrm{M}}$ & 0.592 & 0.731 & 0.832 & 0.980   \\
		BT &  0.316 & 0.342 & 0.190 & 0.303   \\
		CT &  0.271 & 0.309 & 0.395 & 0.617   \\
		GT &  0.162 & 0.249 & 0.302 & 0.516   \\
	\end{tabular} 
	\centering
	\begin{tabular}{cccc}
		\multicolumn{4}{c}{Scale Alternatives ($\sigma^2$)}  \\
		50 & 100 & 500 & 1000  \\
		1.11 & 1.09 & 1.04 & 1.04 \\
		- & - & - & - \\
		0.044 & 0.014 & 0.000 & 0.001   \\
		0.624 & 0.761 & 0.867 & 0.980   \\
		0.604 & 0.747 & 0.863 & 0.972  \\
		0.451 & 0.574 & 0.710 & 0.875   \\
		0.628 & 0.773 & 0.887  & 0.982   \\
		0.055 & 0.050 & 0.029 & 0.014   \\
		0.372 & 0.442 & 0.522 & 0.745   \\
	\end{tabular}
\end{table}

Tables \ref{tab:loc,scal1} and \ref{tab:loc,scal2} show results for multivariate Gaussian distributions with different means or variances. We see that MMD-Pearson has considerably lower power than other tests in all settings. We thus compare the other seven tests in more details. Under  location alternatives, when $d=50$ or 100, MMD-Bootstrap performs admirably, closely followed by  $\textrm{fGPK}_{\textrm{M}}$ and GPK, with fGPK trailing slightly behind.  However, with larger dimensions ($d=500$ or 1000), the new tests surpass MMD-Bootstrap, with $\textrm{fGPK}_{\textrm{M}}$ emerging as the top performer.  
In cases of unbalanced sample design, both GPK and $\textrm{fGPK}_{\textrm{M}}$ exhibit high power. Under scale alternatives, MMD-Bootstrap lags significantly behind the new tests. Among the new tests, GPK and fGPK perform comparably, outperforming $\textrm{fGPK}_{\textrm{M}}$. Notably, BT demonstrates high power for scale alternatives but falls short for location alternatives. 

\begin{table}[h]
	\caption{\label{tab:t}Estimated power of the tests for multivariate $t$-distributed data $(m=n=50)$}
	\centering
	\begin{tabular}{ccccc}
		& \multicolumn{4}{c}{Location Alternatives ($\Delta$)}  \\
		$d$ & 50 & 100 & 500 & 1000   \\
		$\Delta$ $|$ $\sigma^2$ & 0.8 & 1.2 & 1.9 & 2.5 \\
		MMD-Pearson & 0.075 & 0.121 & 0.685 & 0.829   \\
		MMD-Bootstrap & 0.454 & 0.721 &  0.993 & 1.000    \\
		GPK  & 0.397 &  0.690 & 1.000  & 1.000   \\
		fGPK & 0.238 & 0.341 & 0.654 & 0.683    \\
		$\textrm{fGPK}_{\textrm{M}}$ & 0.292 & 0.430 & 0.772 & 0.801  \\
		BT & 0.101 & 0.079 & 0.082  & 0.078   \\
		CT & 0.243 & 0.408 & 0.787 & 0.796  \\
		GT & 0.164 & 0.301 & 0.932 & 0.980  \\
	\end{tabular}
	\centering
	\begin{tabular}{cccc}
		\multicolumn{4}{c}{Scale Alternatives ($\sigma^2$)}  \\
		50 & 100 & 500 & 1000   \\
		1.15 & 1.13 & 1.08 & 1.08 \\
		0.006 & 0.007 & 0.024 & 0.095   \\
		0.131 & 0.248 & 0.249 & 0.564   \\
		0.359 & 0.581 & 0.641  & 0.883   \\
		0.356 & 0.573 &  0.633 &  0.875  \\
		0.380 & 0.613 & 0.677 & 0.900   \\
		0.460 & 0.689 & 0.690  & 0.910  \\
		0.062 & 0.017 & 0.010 & 0.000   \\
		0.272 & 0.376 & 0.292 & 0.408   \\
	\end{tabular}
\end{table}
\begin{table}[h] 
	\caption{\label{tab:chi}Estimated power of the tests for chi-square data $(m=n=50)$}
	\centering
	\begin{tabular}{ccccc}
		& \multicolumn{4}{c}{Location Alternatives ($\Delta$)}  \\
		$d$ & 50 & 100 & 500 & 1000  \\
		$\Delta$ $|$ $\sigma^2$ & 2.05 & 2.90 & 5.36 & 7.90 \\
		MMD-Pearson &  0.072 & 0.043 & 0.006 & 0.011   \\
		MMD-Bootstrap & 0.352 & 0.467 & 0.450 & 0.633   \\
		GPK & 0.330 & 0.437 & 0.738 & 0.988  \\
		fGPK & 0.224 & 0.280 & 0.543 & 0.912   \\
		$\textrm{fGPK}_{\textrm{M}}$ & 0.265 & 0.347 & 0.615 & 0.952 \\
		BT &  0.131 & 0.104 & 0.091 & 0.120   \\
		CT &  0.206 & 0.294 & 0.444 & 0.665   \\
		GT &  0.150 & 0.164 & 0.259 & 0.586   \\
	\end{tabular}
	\centering
	\begin{tabular}{cccc}
		\multicolumn{4}{c}{Scale Alternatives ($\sigma^2$)}  \\
		50 & 100 & 500 & 1000  \\
		1.12 & 1.11  & 1.06 & 1.06 \\
		0.042 & 0.029 & 0.001 & 0.000   \\
		0.247 & 0.369 & 0.068 & 0.013   \\
		0.344  & 0.563 & 0.657 & 0.919  \\
		0.338 & 0.557 & 0.681 & 0.932   \\
		0.375 & 0.605 & 0.698 & 0.939  \\
		0.344 & 0.547 & 0.698  & 0.937   \\
		0.149 & 0.130 & 0.054 & 0.034   \\
		0.193 & 0.272 & 0.372 & 0.565  \\
	\end{tabular}
\end{table}

Table \ref{tab:t} shows results for multivariate $t$-distributed data. We see that MMD-Bootstrap and GPK are very powerful for the mean alternatives and $\textrm{fGPK}_{\textrm{M}}$ also shows good performance. However, MMD-Bootstrap struggles with scale alternatives, whereas the new tests still perform well. CT and GT exhibit high power for  location alternatives but  lose power for  scale alternatives. Conversely, BT exhibits the opposite pattern.

Tables \ref{tab:chi} shows results for chi-square data.  Much like the findings for multivariate Gaussian data, the new tests, especially GPK and $\textrm{fGPK}_{\textrm{M}}$, dominate in power under  location alternatives, especially with larger dimensions  ($d=500$ or 1000). Under scale alternatives, when $d=50$ or 100, $\textrm{fGPK}_{\textrm{M}}$ outperforms the other tests, while BT and fGPK also exhibit high power for larger dimensions ($d=500$ or 1000). These results show that the new tests work well for both symmetric and asymmetric distributions in moderate to high dimensions.

Table \ref{tab:null} shows empirical size of the tests at 0.05 significance level for  multivariate Gaussian and chi-square data. We see that the new tests control the type I error rate well. 
\begin{table}[h] 
	\caption{\label{tab:null}Empirical size of the tests at 0.05 significance level $(m=n=50)$}
	\centering
	\begin{tabular}{ccccc}
		& \multicolumn{4}{c}{Multivariate Gaussian}  \\
		$d$ & 50 & 100 & 500 & 1000   \\
		MMD-Pearson & 0.000 & 0.000 & 0.000 & 0.000   \\
		MMD-Bootstrap & 0.045 & 0.029 & 0.002 & 0.000   \\
		GPK  & 0.044 & 0.051 & 0.048 & 0.046   \\
		fGPK & 0.042 & 0.038 & 0.041 & 0.043   \\
		$\textrm{fGPK}_{\textrm{M}}$ & 0.047 & 0.043 & 0.056 & 0.054   \\
		BT &  0.043 & 0.047 & 0.050 & 0.047   \\
		CT &  0.054 & 0.055 & 0.075 & 0.059   \\
		GT &  0.045 & 0.053 & 0.048 & 0.041   \\
	\end{tabular}
	\centering
	\begin{tabular}{cccc}
		\multicolumn{4}{c}{Chi-square}  \\
		50 & 100 & 500 & 1000   \\
		0.002 & 0.000 & 0.000 & 0.000   \\
		0.042 & 0.022 & 0.002 & 0.000  \\
		0.046 & 0.040 & 0.044  & 0.054   \\
		0.042 & 0.025 & 0.038 & 0.044  \\
		0.048 & 0.039 & 0.050 & 0.055   \\
		0.049 & 0.046 & 0.050  & 0.055   \\
		0.055 & 0.056 & 0.044 & 0.058   \\
		0.045 & 0.052 & 0.045 & 0.044   \\
	\end{tabular}
\end{table}

Overall, the power tables illustrate that the new tests perform well across a wide range of alternatives.  GPK demonstrates strong performance across all settings, while fGPK maintains high power with the added advantage of computational efficiency.  
Unlike MMD tests, $\textrm{fGPK}_{\textrm{M}}$ is both computationally efficient and capable of capturing  variance differences to some extent. In practice, fGPK and $\textrm{fGPK}_{\textrm{M}}$ are preferred due to their speed and effectiveness across a broad spectrum of alternatives.  If further investigation is required, GPK could be useful.


\section{Real Data Examples} \label{sec:data}

\subsection{Musk data}

We first illustrate the new tests on the Musk dataset \citep{blake1998uci}, which is publicly accessible at \url{https://archive.ics.uci.edu/ml/datasets.php}. The  dataset comprises molecule structure data where the features represent the shape of the molecules constructed by the rotation of bonds. It contains information about 476 molecules (each has dimension $d=166$), with 269 classified as musks by human experts and the remaining 207 classified as non-musks. 
We employ this dataset to illustrate how the new tests distinguish musks versus non-musks based on molecular shape. Here, we conduct the tests on subsets of the entire dataset to compare their empirical power. For each value of $m$, we randomly draw $m$ observations from the pool of 269 musk observations and $m$ observations from the pool of 207 non-musk observations. The process is repeated  1,000 times, and  the significance level of the test is set to be 0.01.  Table \ref{tab:musk} displays the estimated power of the tests for various values of $m$. The results indicate that the new tests consistently outperform MMD-Person and MMD-Bootstrap across different sample sizes, demonstrating the consistent enhancement provided by the new testing methods.

\begin{table}[h]
	\caption{\label{tab:musk}Estimated power of the tests}
	\centering
	\begin{tabular}{cccccc}
		$m$ & 30 & 40 & 50 & 60 & 70 \\
		MMD-Pearson  &  0.058 &   0.121 &   0.190 & 0.270 & 0.402 \\
		MMD-Bootstrap & 0.091 & 0.167 & 0.275 & 0.388 & 0.568 \\
		GPK &  0.133 & 0.265 &  0.434 & 0.606 &  0.780  \\
		$\textrm{fGPK}$ &  0.260 & 0.445 &  0.618 & 0.742 &  0.865  \\
		$\textrm{fGPK}_{\textrm{M}}$ &  0.077 & 0.215 &  0.301 & 0.437 &  0.639  \\
	\end{tabular}
\end{table}

\subsection{New York City taxi data}

We here illustrate the new tests on the New York City taxi dataset, which is publicly accessible on the New York City Taxi \& Limousine Commission website (\url{https://www1.nyc.gov/site/tlc/about/tlc-trip-record-data.page}). This dataset contains various details, including latitude and longitude coordinates of pickup and drop-off locations,  pickup and drop-off date, driver-reported passenger counts, fares, and more.  Given the richness of the data, we utilize it to illustrate the new tests by examining travel patterns over consecutive months. 

\begin{figure}[h]
	\centering
	\subfloat[January 1, 2015]{\includegraphics[width=2.5in]{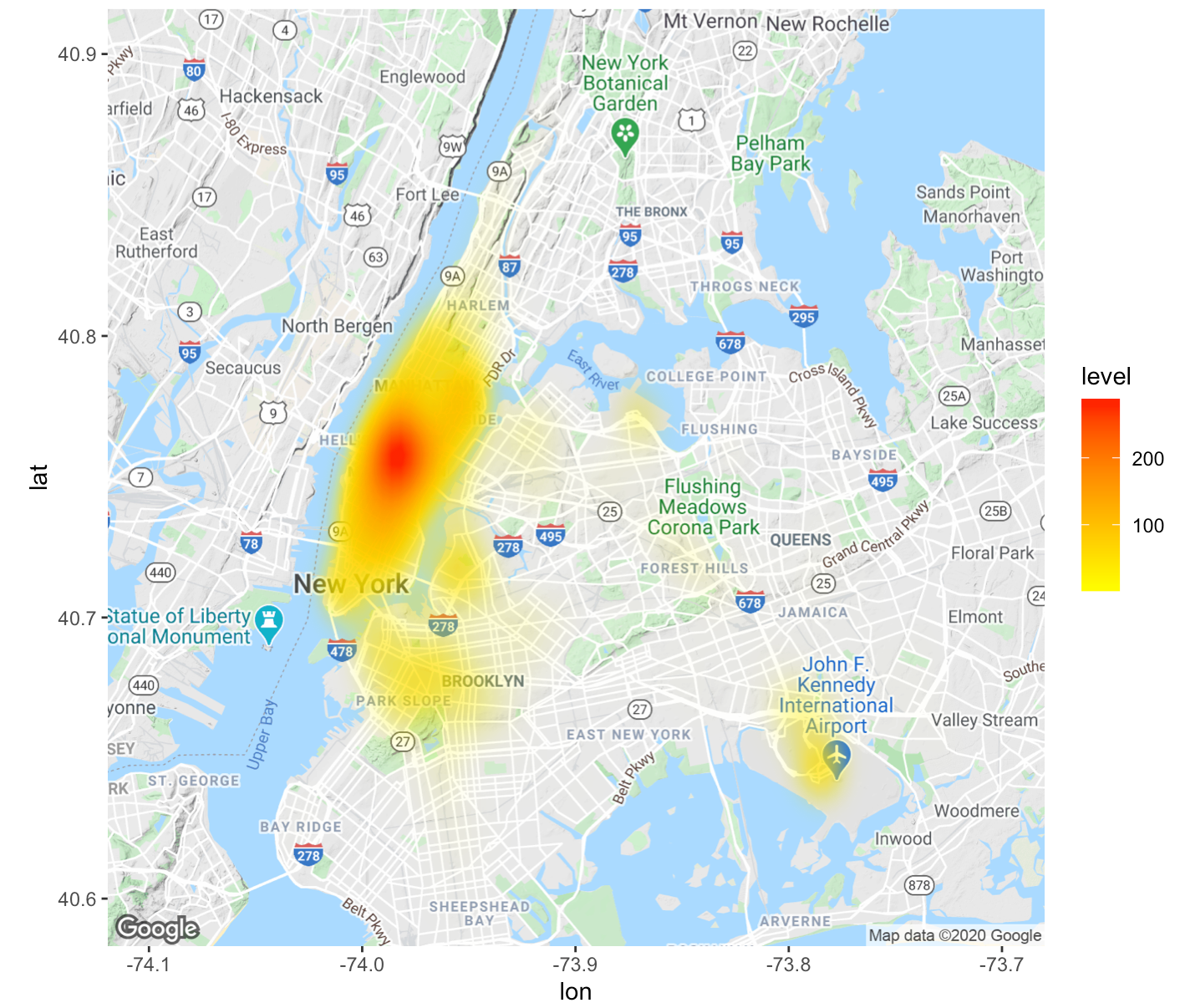}} \ \ 
	\subfloat[February 1, 2015]{\includegraphics[width=2.5in]{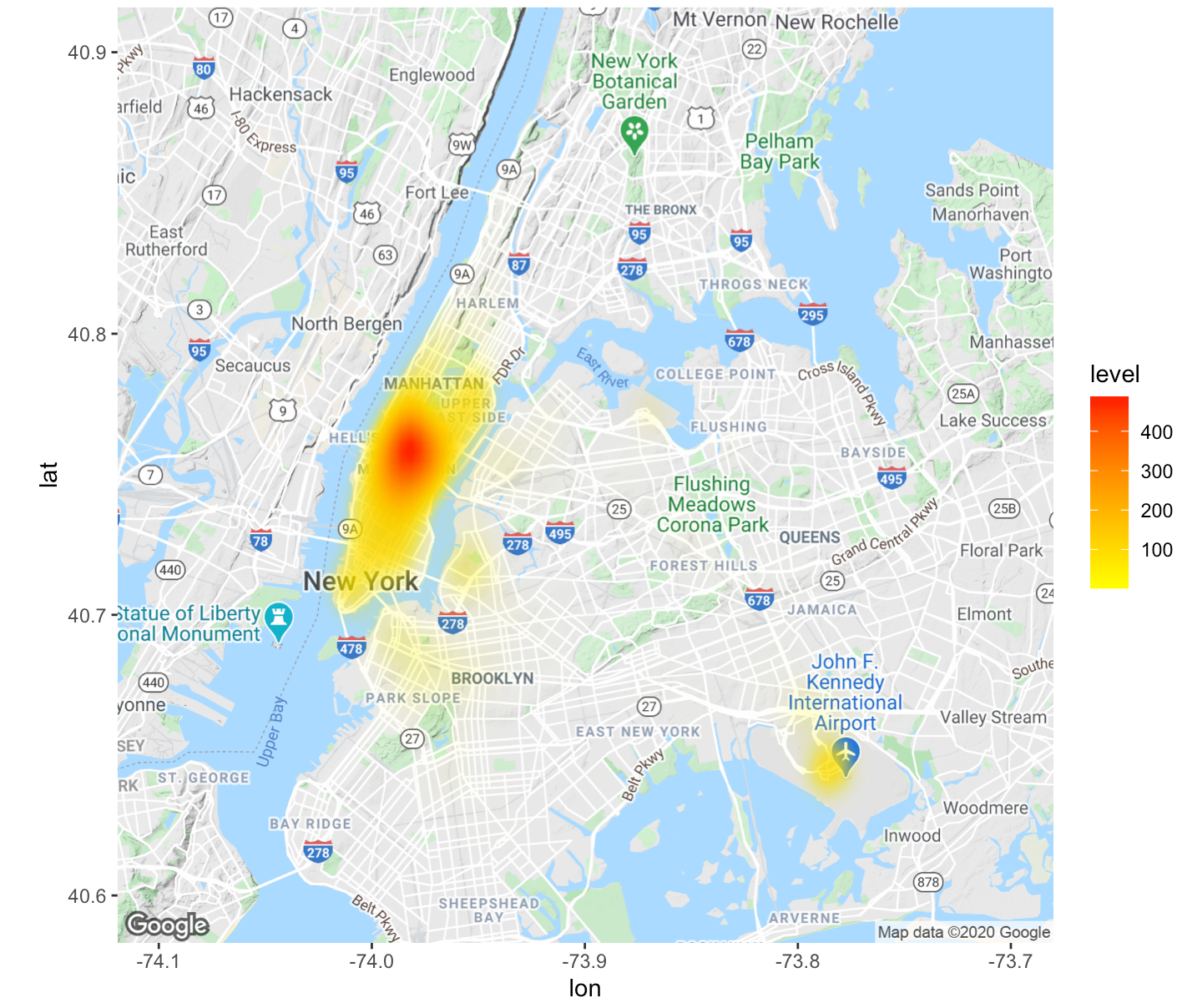}}
	\caption{Density heatmaps of taxi drop-off locations of trips originating from John F.~Kennedy airport on January 1st and February 1st in 2015.}
	\label{airport}
\end{figure}

Specifically, we focus on trips originating from John F.~Kennedy international airport. We preprocess the data similarly to \cite{chu2019asymptotic}, defining the John F.~Kennedy airport boundary with latitude coordinates ranging from  40.63 to 40.66  and longitude coordinates ranging from -73.80 to -73.77. Figure \ref{airport} provides density heatmaps representing the drop-off locations of  trips originating from John F.~Kennedy airport on two distinct days, January 1st and February 1st in 2015. We divide this area into a 30$\times$30 grid with equal cell sizes and count the number of trips whose drop-off locations fall within each cell for each day. Then, we use these 30$\times$30 matrices to access whether there are differences in travel patterns between January and February in 2015. To accomplish this, we measure the distance of two matrices as the Frobenius norm of their difference of the two matrices, and use the Gaussian kernel with the median of all pairwise distances as the bandwidth for all kernel tests.

Table \ref{tab:taxi2} shows the results of the tests. Notice that MMD-Pearson cannot be applied due to unbalanced sample sizes. We see that the new tests reject the null hypothesis of equal distributions at 0.05 significance level, while MMD-Bootstrap does not.

\begin{table}[h]
	\caption{\label{tab:taxi2}$p$-values of the tests}
	\centering
	\begin{tabular}{ccccc}
		& MMD-Bootstrap  & GPK & fGPK & $\textrm{fGPK}_{\textrm{M}}$ \\
		Jan vs. Feb & 0.141 & 0.031 &0.008 &0.005 \\
	\end{tabular}
\end{table}

We delve deeper into the test statistics for this comparison. Table \ref{tab:taxi3} displays the $\alpha-\gamma$ and $\beta-\gamma$ values and their standardized values, as well as the $p$-values of the test based on $Z_{W,1.2}$, $Z_{W,0.8}$, and $Z_{D}$.  We see that $\alpha-\gamma$ is negative and offsets with $\beta-\gamma$, causing MMD-Bootstrap to be insignificant.  When examined separately, the standardized values of $\alpha-\gamma$ and $\beta-\gamma$ are relatively large, indicating a significant variance difference.  The relatively small $p_{D}$ and the vary small $p_{W,0.8}$, which encompasses this specific alternative, contribute to GPK, fGPK, and $\textrm{fGPK}_{\textrm{M}}$ capturing the difference effectively. 

\begin{table}[h]
	\caption{\label{tab:taxi3} Breakdown values, $(\alpha-\gamma)^* = \frac{\alpha-\gamma - \ep(\alpha-\gamma)}{\surd{\var(\alpha-\gamma)}}$, $(\beta-\gamma)^* = \frac{\beta-\gamma - \ep(\beta-\gamma)}{\surd{\var(\beta-\gamma)}}$} 
	\centering
	\begin{tabular}{ccccccccc}
		Jan vs. Feb & $\alpha-\gamma$ & $\beta-\gamma$ & $(\alpha-\gamma)^*$ & $(\beta-\gamma)^*$ & MMD & $Z_{W,1.2}$ & $Z_{W,0.8}$ & $Z_{D}$ \\
		Value & -0.061 &0.070 & -2.35 & 2.71 & 0.009 & -1.164 & 2.781 & -2.547 \\
		$p$-value & - & - &  - & - & - & 0.88 & 0.0027 & 0.011 \\
	\end{tabular}
\end{table}





\section*{Acknowledgement}
Hoseung Song and Hao Chen were supported in part by the NSF award DMS-1848579.  The authors thank Yejiong Zhu for insightful discussions on asymptotic properties of the statistics.


\section*{Supplementary material}

Supplementary material available at \emph{Biometrika} online contains proofs for Theorem 1--6, an illustration on conditions in Theorem 2, runtimes of the fast tests implemented in $\texttt{R}$, and additional simulation results.


\bibliographystyle{biometrika}
\bibliography{Bibliography-MM-MC}

\end{document}